\documentclass[final,3p,times,twocolumn]{myelsarticle}
\usepackage{amssymb,amsmath}
\def\sfrac#1#2{{\textstyle{#1\over #2}}}
\newcommand{\be}{\begin{equation}}
\newcommand{\ee}{\end{equation}}
\newcommand{\ba}{\begin{array}}
\newcommand{\ea}{\end{array}}
\newcommand{\bea}{\begin{eqnarray}}
\newcommand{\eea}{\end{eqnarray}}

\newcommand{\bfpi}{{\boldsymbol\pi}}
\newcommand{\bfS}{{\boldsymbol S}}

\journal{Physics Letters B, vol. 173, number 2, pp.\ 173-178}

\begin{document}

\begin{frontmatter}



\title{THE VALIDITY OF PERTURBATION THEORY
FOR THE $O(N)$ NON-LINEAR SIGMA MODELS\tnoteref{l1}}
\tnotetext[l1]{Work supported in part by US Department of Energy under
contract DEAC-03-81-ER40050}


\author{James M. Cline}

\address{California Institute of Technology, Pasadena, CA 91125, USA
\\ \bigskip Received 14 March 1986}

\begin{abstract}
Recently it has been claimed that ordinary perturbation theory (OPT) gives incorrect weak coupling expansions for lattice
$O(N)$ non-linear sigma models in the infinite volume limit, and in particular that the two-dimensional non-abelian models are
not asymptotically free, contrary to previous findings. Here it is argued that the problem occurs only for one-dimensional
infinite lattices, and that in general, OPT gives correct expansions if physical quantities are first computed on a finite lattice,
and the infinite volume limit is taken at the end. In one dimension the expansion is sensitive to boundary conditions because of
the severe infrared behavior, but this is not expected to happen in higher dimensions. It is concluded that spin configurations
which are far from the perturbative vacuum have too small a measure in the path integral to invalidate OPT, even though they
are energetically allowed for non-zero values of the coupling.

\end{abstract}



\end{frontmatter}

Two-dimensional spin systems have been important
field theoretic laboratories because of their similarities
to four-dimensional gauge theories. For example,
when $N > 2$ the $O(N)$ symmetry is non-abelian
and the theory has been shown to be asymptotically
free [1,2]. The sigma model could also be regarded
as a testing ground for perturbation theory. As for
gauge theories, the perturbative vacuum, with all
spins aligned, is quite different from the true vacuum,
since there is no magnetization for $d \le 2$ [3].
A further similarity is that the expansion proceeds
in powers of the coupling {\it and} the fields, although
there is no mass to damp out configurations far away
from the perturbative vacuum. Therefore one might
worry that very long wavelength excitations, where
the fields eventually get large, will contribute significantly
to the path integral, and that ordinary perturbation
theory (OPT) may not account for them
correctly.

Thusly Patrascioiu has argued that OPT gives incorrect low-temperature
expansions for the free energy and spin correlations of the $O(N)$
non-linear sigma model, for $N> 2$ and infinite volume lattices [4]. 
(For
finite lattices there always exist temperatures low enough so that the
spins are all relatively aligned, and OPT should have no problem.) To
remedy this, he formulates a new perturbation expansion which treats
the gradients of the fields, rather than the fields themselves, as
small quantities. Despite the method's computational difficulty (it is
very nonlocal), he extracts a value for the one-oop spin-spin
correlation function in two dimensions, which yields an opposite sign
for the Callan-Symanzik $\beta$ function, relative to the OPT result. 

It is
important to investigate this claim, for if it is true, then it could
conceivably be that the $\beta$ function for QCD or other gauge theories is
different---in magnitude, if not in sign---from what is presently
accepted. In this letter I show that OPT gives the correct result in
one dimension, where the theory is exactly soluble, {\it if it is
formulated on a finite lattice with the correct boundary condition}.
Since this result is independent of the lattice size, $L$, it is
trivially correct in the infinite volume limit. This procedure should
give correct results in higher dimensions, where the IR behavior is
milder. It is shown that starting with an infinite lattice gives
results consistent with the exact solution for $N \to\infty$ (as well as the
infinite volume limit of the finite lattice calculations) except
in the one-dimensional models, apparently because
of their more severe IR divergences.

The controversy over OPT's validity stems from
the difference between the OPT result for the free
energy density on a $d$-dimensional, infinite lattice [5],
\bea
	{\ln Z\over L^d} = {d\over g} + \sfrac12 (N-1)\ln g 
+	{(N-2)\over 8d}g + O(g^2)
\eea 
versus the expansion of the exact result for d = 1 [6]
\bea
	{\ln Z\over L} = {1\over g} + \sfrac12 (N-1)\ln g 
	-\sfrac18{(N-1)(N-3)}g \nonumber\\
+ O(g^2)
\eea 
where the lattice partition function is given by
\be
Z = \int\prod_x{\rm d}{\bf S}\,\delta({\bf S}^2_x-1)\exp\left(
{1\over g}\sum_x\sum_{\mu=1}^d{\bf S}_x\cdot{\bf
S}_{x+\hat\mu}\right).
\ee
OPT proceeds by rewriting ${\bf S}$ as $(\bfpi, \sigma)$, 
where ${\bfpi}$ is
$(N-1)$-dimensional, and solving the constraint, so
that $\sigma = \pm(1-{\bfpi}^2)^{1/2}$. The vacuum state is chosen to
be $\sigma = 1$, and configurations with $\sigma < 0$ are discarded,
as they give, naively, contributions of $O(\exp(-1/g))$.
After rescaling $\bfpi^2$ to $g\bfpi^2$, $Z$ becomes
\bea
&&\phantom{.} \!\!\!\!\!\!\!\!\!\!\! Z =\exp(d L^d/g)\int\prod_x{{\rm
d}\bfpi_x\over(1-g\bfpi_x^2)^{1/2}}\ \times\\
&&\phantom{.} \!\!\!\!\!\!\!\!\!\!\!\exp\left(-\sum_{x,\mu}\sfrac12(\Delta_\mu\bfpi_x)^2-{1\over 2g}
\sum_x\left[\Delta_\mu(1-g\bfpi^2)^{1/2}\right]^2\right).\nonumber
\eea
To do perturbation theory, the integration region
must be extended from $\bfpi^2 \le 1/g$ to $\bfpi^2 \le \infty$ and the
radicals Taylor-expanded.

More crucially, the zero modes of the $\bfpi$ field must
be removed. This could be accomplished by introducing
a magnetic field in the ${\boldsymbol e}_N$ direction 
(i.e., $\bfpi = 0$ and
$\sigma = 1$), which is equivalent to a mass for the $\bfpi$,
 and removing
it at the end of the calculation. However, this
method is known to give wrong results even for a two-spin
system, beyond the tree level [7]. It is easy to
see why this happens, heuristically: when the factor
$(1 -g\bfpi^2)^{1/2}$ in the measure is reexpressed as a term
in the action, it contributes a mass of the wrong sign,
$m^2 = -g$. Then, as the magnetic field becomes smaller
than $g$, the action becomes unbounded from below,
and the gaussian integrations no longer make sense.
A better way is to use the global $O(N)$ invariance to
fix one of the spins, say $\bfS_0$, to be in the ${\boldsymbol e}_N$
 direction. More generally, one can use the Faddeev-Popov procedure
to fix any linear combination of the $\bfS_x$, thus
removing the zero modes, and this gives agreement
with exact solutions in every case that has been
checked [7]. However, fixing spins is not sufficient
for an infinite lattice, because it provides a low-momentum
cutoff of $O(1/L)$ ($L$ being the lattice size)
which vanishes as $L \to \infty$ causing the $\bfpi$ propagators
to become undefined. Thus one is forced to bring in
a magnetic field again. Since this is a bad procedure
for finite systems, there is no a priori reason for it to
work on an infinite lattice. But this is exactly how
(1) was obtained; hence the discrepancy with (2).

\begin{figure*}[t]
\centerline{\includegraphics[width=0.65\textwidth]{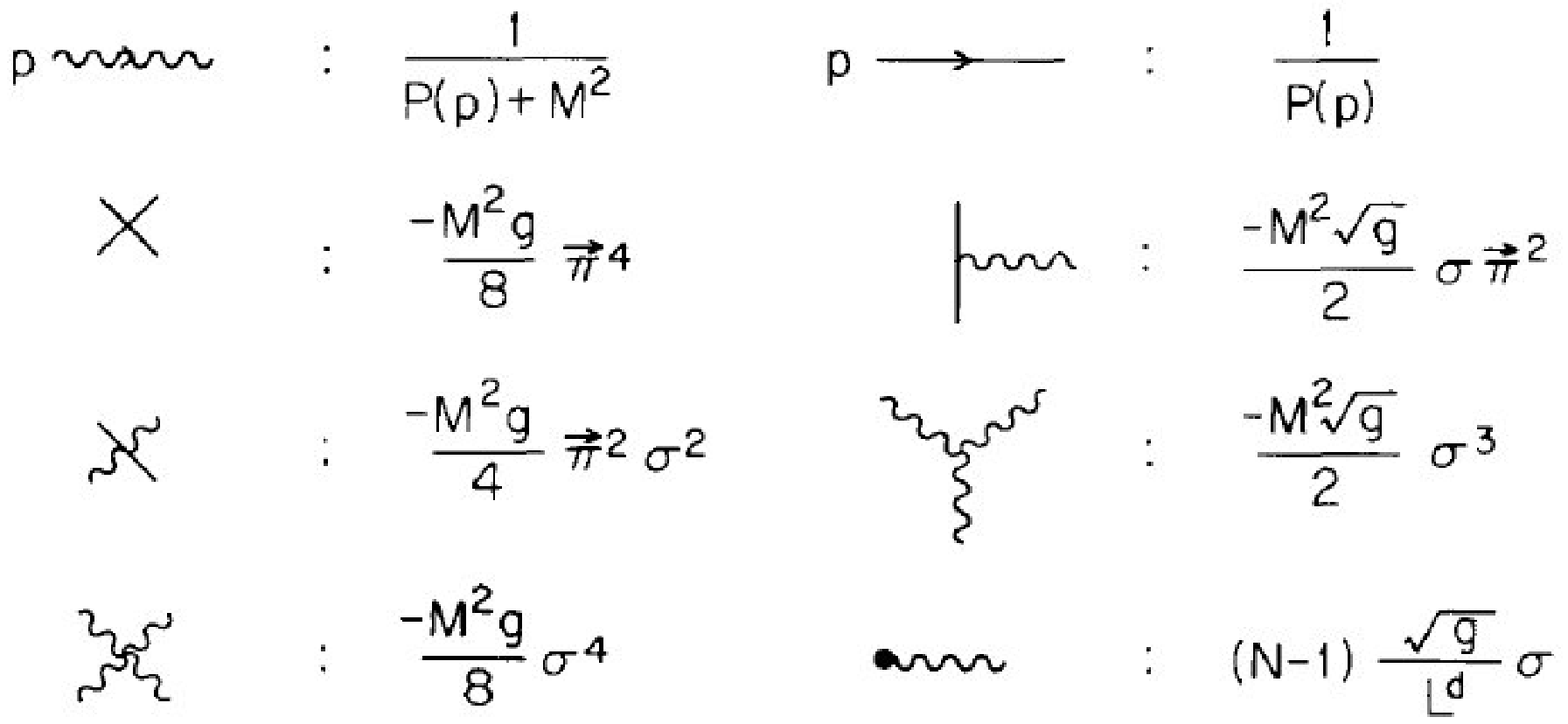}}
  \caption{Propagators and vertices for the linear $\sigma$ model, 
including the lowest order vertex from the Faddeev-Popov determinant}
\end{figure*}

Patrascioiu has a different explanation for why
OPT goes wrong as $L\to\infty$ in 1 dimension, however.
Note that the two-point function is given by
\bea
\langle\bfS_0\cdot\bfS_x\rangle &=& g\langle\bfpi_0\cdot\bfpi_x\rangle
\nonumber\\
&&
+\langle(1-g\bfpi^2_0)^{1/2}(1-g\bfpi^2_x)^{1/2}\rangle\nonumber\\
&=&1 + g(N-1)D(x)+O(g^2)
\eea
at tree level, where $D(x) \equiv G(x) - G(0)$ and $G(x)$ is
the massless scalar propagator in $d$ dimensions. The
large-$x$ dependence of $D(x)$ is $-|x|$, $-\ln|x|$ and $-|x|^{-1}$
for  $d = 1,2,3$, respectively. Therefore we
have configurations in which the spins wander far
away from ${\boldsymbol e}_N$, the perturbative vacuum, 
over a distance
$|x|\sim g^{-1}$ ($|x|\sim \exp(g^{-1})$) for $d = 1$ ($d = 2$),
whereas these are energetically suppressed for $d \ge 3$
(magnetization). Of course, OPT is not designed to
account for spins pointing far away from ${\boldsymbol e}_N$, since
$\sigma < 0$ was discounted. The argument is that for finite
$L$ there should be no problem, because there always
exists a $g$ small enough so that $gL$ or $g\ln L$ is $\ll 1$:
then no large excursions away from ${\boldsymbol e}_N$ are allowed.
For $L =\infty$, $d \le 2$, the phase transition occurs exactly
at $g = 0$, so that for finite $g$ these long spin waves are
unsuppressed, and OPT may fail.

In order to clarify why OPT fails in one dimension,
I compute the correlation function $\langle\bfS_0\cdot\bfS_x\rangle$
 in $d$ dimensions,
using a somewhat different method, which
is to obtain the non-linear $\sigma$ model from the infinite
mass limit of the linear model. This has the advantage
of avoiding the OPT approximations which made the
long spin waves impossible to represent correctly. It
is instructive to do the calculation on both finite and
infinite lattices, for it will become apparent that only
in one dimension does any discrepancy appear.

First consider the finite lattice. We start by noticing
that the $\delta$ functions in (3) can be rewritten using
the identity $\delta(x) = \lim_{M\to\infty}
(M/\sqrt{\pi})\exp(-M^2x^2)$.
After the change of variables $\bfS \to (\sqrt{g}\bfpi, 1 +
\sqrt{g}\sigma$) and
``gauge fixing,'' the partition function is
\bea
\phantom{.}\!\!\!\!\!\!\!Z&=&\lim_{M\to\infty}\left(M\over\sqrt{8\pi}\right)^{L^d}
	\exp\left[L^d\left({d\over g}+\sfrac12(N-1)\ln g\right)\right]
	\nonumber\\
&\times&\int{\rm D}\bfpi\,{\rm D}\sigma\,\left[\Delta_{F}(\bfpi)\,
	\delta(F(\bfpi))\right]\\
&\times&\exp\left(-\sfrac12\sum_{x\mu}\left[(\Delta_\mu\bfpi_x)^2 +
(\Delta_\mu\sigma_x)^2 + M^2\sigma_x^2\right]\right.\nonumber\\
&-&\left.M^2\sum_x\left[\sfrac18 g(\bfpi_x^2+\sigma_x^2)^2+
	\sfrac12\sqrt{g}\sigma_x(\bfpi_x^2+\sigma_x^2)\right]\right)
\nonumber
\eea
where $\delta(F)$ removes the zero modes of $\bfpi$, and $\Delta_F$ is
the associated determinant. A convenient choice
which preserves the $\bfpi$ propagator's translational invariance
is $F = \sum_x \bfpi_x$, in which case $\Delta_F$  turns out to
be $(1 + \sqrt{g}L^{-d}\sum_x\sigma_x)^{N-1}$ (see ref. [7]), and the
bare propagator is
\bea
\langle\bfpi_x\cdot\bfpi_0\rangle^{(0)} &=& (N-1)\, G(x)\nonumber\\
&=& (N-1)\,L^d\sum_{p\ne 0}{\exp(i2\pi p\!\cdot\! x/L)\over
P(p)},\nonumber\\
P(p) &=& 4\sum_\mu\sin^2\pi p_\mu/L\, .
\eea

\begin{figure*}[t]
\centerline{\includegraphics[width=0.75\textwidth]{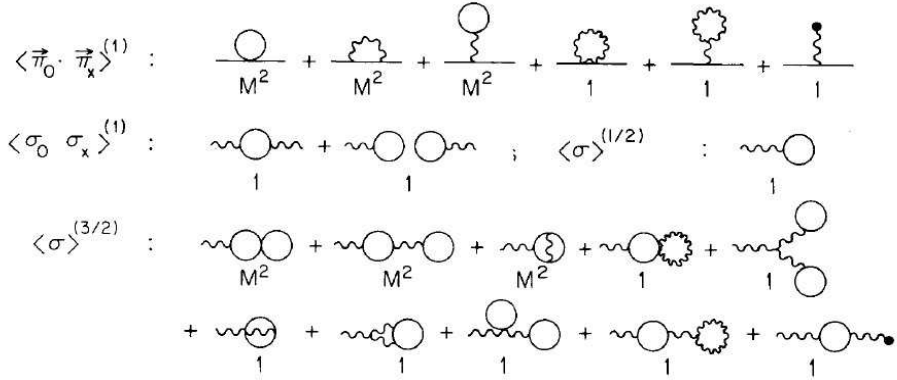}}
  \caption{Diagrams contributing to $\langle\bfS_0\cdot\bfS_x\rangle$
to second order, with leading $M$ dependence greater than $M^{-2}$.}
\end{figure*}

The propagators and vertices of this theory are shown
in fig.\ 1, including the lowest order vertex due to $\Delta_F$.
To compute the two-point function
\bea
\langle\bfS_0\cdot\bfS_x\rangle &=& 1 + \sqrt{g}
	(\langle\sigma_x\rangle
	+\langle\sigma_0\rangle)\nonumber\\
	&+&
g(\langle\sigma_0\sigma_x\rangle+\langle\bfpi_0\cdot\bfpi_x\rangle)
\eea
to $O(g^2)$, we need the diagrams listed in fig.\ 2, whose
leading $M^2$-dependences are nonvanishing. Fig.\ 3 gives
examples of graphs which vanish as $M \to\infty$ because
they have more $\sigma$ propagators than non-$\Delta_F$ vertices.
Although some of the graphs in fig.\ 1 diverge like $M^2$,
their sums are finite, as expected, and the factor $M^{L^d}$
in $Z$ is exactly cancelled by $[\det(-\Delta^2 +M^2)]^{-1}$ coming
from the $\sigma$ integrations. After much algebra, the
one-loop contribution to $\langle\bfS_0\cdot\bfS_x\rangle$ obtained 
is
\bea
\phantom{.}\!\!\!\!\!\!\!\langle\bfS_0\cdot\bfS_x\rangle^{(2)}&=&
g^2\Bigg(-(N-1)(N-2)\\
&\times&{1\over L^{2d}}\sum_{p\ne 0}{\exp(i2\pi p\!\cdot\! x/L)-1\over
P^2(p)}\nonumber\\
&+&\!\!\!\!\!\!\sfrac12(N-1)D(x)\left[{1\over d}(1-L^{-d})+D(x)\right]\Bigg)
\nonumber
\eea
in exact agreement with the OPT result of Hasenfratz
[7]. This just shows that the OPT approximations---ignoring 
$\sigma < 0$ configurations, and enlarging the integration
region beyond the radius of convergence of
$(1 - g\bfpi^2)^{1/2}$'s Taylor series---are justified for $L <\infty$
as expected.

Before carrying out the finite lattice calculation
of $\langle\bfS_0\cdot\bfS_x\rangle$, 
I will show that the procedure leading
to (9) is correct for $d = 1$ and any $L$ (and thus, as
$L \to\infty$). This case is important because: (1) only here
is the exact solution known (unless $N = \infty$); (2) the
low-energy spin waves should affect OPT most severely
when $d = 1$, by Patrascioiu's argument; and (3) the
only known failure of OPT occurs when $d = 1$. For
simplicity, let us only consider the average interaction
energy per spin,
\be
	f={1\over L}\sum_n\langle\bfS_n\cdot\bfS_{n+1}\rangle =
	-{g^2\over L}{\partial\over\partial g}\ln Z
\ee
From (9) we find that
\bea
	\lim_{L\to\infty}f_{OPT} &=& 1 - \sfrac12
g(N-1)\nonumber\\
	&+&\sfrac1{24} g^2 (N-1)(N-5) + O(g^3)
\eea
whereas the expansion of the exact result is [6]
\bea
	f_{\rm exact} &=& 1 - \sfrac12 g(N-1)\nonumber\\
	&+&\sfrac18 g^2 (N-1)(N-3)  + O(g^3)
\eea
independent of $L$. In comparing these it must be remembered
that the exact result is obtained for an
open chain of spins, in which the ends do not interact
with each other, whereas OPT is normally done
on a periodic lattice, which is the case in (11). To
repeat the OPT calculation (4) for the open chain,
one must use the open chain lattice propagator,
which in $\sum_x \bfpi_x = 0$ gauge and at $O(g^0)$ is exactly
\be
G_{\rm open}(x,y) = G_{x,y}+L\,(G_{xL} -G_{x1})(G_{yL} -G_{y1}).
\ee
Here $G_{xy}$ is the periodic lattice propagator in (7),
which for $d = 1$ has the closed form
\bea
G_{xy} &=& \sfrac1{12}(L-L^{-1})-\sfrac12|x-y{\rm\ mod}\ L|\nonumber\\
	&+&{1\over 2L}(x-y{\rm\ mod}\ L)^2
\eea
Therefore the momentum sums in loop graphs can
be done explicitly. The result is (12), including the
exact $L$-independence which distinguished it. {\it Thus,
OPT agrees with the exact result in $d = 1$ as long as
the same boundary conditions as in the exact solution
are used.} This does not mean that (11) is incorrect.
To prove that the difference between (11) and
(12) is due to the physical difference made by the
extra link interaction, and not some fluke of OPT,
I applied Patrascioiu's method to the periodic spin
chain with $L$ spins and confirmed (11). (This letter
has no complaint with the method of ref.\ [4], but
only its result.)

\begin{figure}[b]
\centerline{\includegraphics[width=0.25\textwidth]{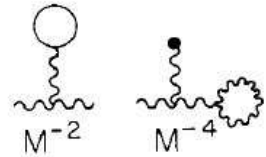}}
  \caption{Examples of graphs which vanish as $M\to\infty$.}
\end{figure}

Of course, the exact expressions for $f$ must agree
for the two boundary conditions as $L\to\infty$. The only
plausible explanation for the difference in the perturbative
expansions is that the exact expressions
differ by a function like
\be
f_{\rm open} - f_{\rm closed}\sim 
L^{-1} \exp[-g^2L(N- 1)(N- 2)/12]
\ee
for example. Even though this vanishes as $L\to\infty$, its
expansion in powers of $g$ does not, and in fact it predicts
that the weak-coupling expansion of $f_{\rm closed}$ is
IR-divergent, starting in $O(g^4)$. From the point of
view of the perturbative calculation, the reason for
sensitivity to the boundary condition, which is a $1/L$
effect, is that the propagator is linear in distance and
so acquires values of $O(L)$ in the loop diagrams. In
two dimensions the propagator is logarithmic, so we
might expect the $L$ in the exponent of (15) to be replaced
by $\ln L$. The resulting expression cannot lead
to differences in perturbation expansions, so the effect
explained by (15) is almost certainly unique to
$d=1$.

Having seen that OPT on a finite lattice gives the
correct expansion for the energy density as $L\to\infty$, in
one dimension, we now examine what happens when
$L$ is taken to be infinite at the outset. In this case, as
noted previously, the Faddeev-Popov terms in (6)
are insufficient for regulating the $\bfpi$ propagator; the
field must be given a mass, $\mu$, to be eventually removed.
But this spoils the $O(N)$ symmetry, and so
one is not justified to insert gauge-fixing terms:
$\Delta_F\,\delta(F)$ should no longer appear in (6). (Actually,
for the gauge choice $F = \sum_x\bfpi_x$ it turns out not to
matter whether one keeps the $\Delta_F\,\delta(F)$ factor, when
$L =\infty$.) It is straightforward to show that the only effect
of these changes on the result, (9), are (1) to replace
the second term by its $L\to\infty$ limit (recall that
$D(x)$ is IR-finite); and (2) to replace the first term,
the $\Delta_F$ contribution, by $-\sfrac18 g^2(N - 1)|x|^2$ (zero)
when $d = 1$ ($d > 1$). This agrees fully with the OPT
calculation on an infinite lattice for $d = 2$ which was
done by Elitzur [2]. Since the present method was
designed to correctly measure the contributions of
the long spin waves, yet it agrees with OPT for $d\ge 2$,
{\it we conclude that the long spin waves do not cause
OPT to give incorrect results when $d \ge 2$}. Furthermore,
the problem in $d = 1$ is clearly seen to be due
to the noncommutativity of the two limits $g\to 0$
and $\mu\to 0$ (where by $g\to 0$ I mean developing the
asymptotic expansion), just as was suggested earlier.
On the other hand, this analysis shows that these
limits do commute for $d\ge 2$. (More precisely, the
order $\mu\to 0$, $g\to 0$, $L\to\infty$ commutes with $L\to\infty$,
$g\to 0$, $\mu\to 0$ for $d\ge 2$.)

This conclusion derives further support from comparison
with the $N =\infty$ limit of the model, where
again exact solutions are known [8]. This limit exists
if the coupling is rescaled so that $\beta\equiv 1/gN$ is held
fixed as $N\to\infty$. Then, for example, (12) would become
\be
	f_{\stackrel{\mbox{$\scriptstyle N\to\infty$}}{d=1}} = 1-{1\over 2\beta} + {1\over
8\beta^2} + O(\beta^{-3})
\ee
and using the fact that $D(1) = -\sfrac14$ for $d = 2$, (5) and
(9) give us
\be
	f_{\stackrel{\mbox{$\scriptstyle N\to\infty$}}{d=2}} = 
	\langle\bfS_0\cdot\bfS_1\rangle = 1-{1\over 4\beta} + O(\beta^{-3})
\ee
There should be no problem with first doing the small
$g$ expansion and then taking the $L\to\infty$ limit, since
the latter requires $g$ to be infinitesimal. Indeed, the
exact solution for $N = \infty$, in one dimension [6] is
\be
	f_{\stackrel{\mbox{$\scriptstyle N\to\infty$}}{d=1}} = 
	{2\beta\over  1 + (1+4\beta^2)^{1/2}}
\ee
which agrees with (16) when expanded in powers of
$1/\beta$. In two dimensions, the exact solution [9] is
\bea
	f_{\stackrel{\mbox{$\scriptstyle N\to\infty$}}{d=2}} &=& 
	\langle\bfS_0\cdot\bfS_1\rangle = {1-\sfrac14
m^2\over\beta}\times\\
&&\!\!\!\!\!\!\!\!\!\!\int{{\rm d}^2p\over(2\pi)^2}\, 
{\sfrac12 \sum_\mu\cos p_\mu\over 4\sum_\mu\sin^2\sfrac12 p_\mu
	+\sfrac12 m^2\sum_\mu\cos p_\mu}\nonumber
\eea
where $m$ is a dynamically generated mass, which for
large $\beta$ is given by
\be
	m^2 = 32\,\exp(-4\pi\beta) +O(\exp(-8\pi\beta))
\ee
For large $\beta$, hence small $m$, the integral is dominated
by its $p\cong 0$ contributions, and we can approximate
it by
\bea
	&&\!\!\!\!\!\!\!\!\!\!\!\!\!\!\!\!\!{1\over 2\pi}\int_0^\Lambda{\rm d}p\, {p\over p^2+m^2}
	\cong{1\over 4\pi}\ln(\Lambda^2/m^2)\nonumber\\
	&=& \beta - {1\over 4\pi}\ln(32/m^2) + O(\exp(-4\pi\beta)),
\eea
where $\Lambda = O(1)$.  The important thing to notice is
that when this is substituted into (19), the expansion
of $f$ in powers of $1/\beta$ terminates after the $1/\beta$ term.
{\it This agrees with the OPT result, (17), whose $1/\beta^2$
term is seen to vanish.} In contrast, Patrascioiu's result
for the energy density in $d = 2$ contains a piece of order
$g^2 N^2$ which when reexpressed in the large-$N$ limit
is a $1/\beta^2$ term. There seems to be no subtlety of ordering
limits in this comparison, since (19) was derived
for an infinite lattice, and as has already been
mentioned, $g \to 0$ and $N \to\infty$ are expected to commute.

Finally, the $\beta$ function calculation of Polyakov  [1]
merits attention because it is on a completely independent
footing from the OPT method, yet it gets the
same answer, $\beta(g) = -g^2(N-2)/2\pi + O(g^2)$. The
salient point is that it uses Wilson's method of integrating
out the high-momentum components of the
spin field and seeing how the coupling is renormalized
in the effective action for the remaining low-momentum
components. Thus, $\beta(g)$ is determined by the
short wavelength fluctuations, and is insensitive to
the long spin wave effects by which Patrascioiu explains
the difference between his $\beta$ function and the
standard result.

In conclusion, I have shown that OPT on a finite
lattice is capable of giving the correct asymptotic
weak-coupling expansion for $d = 1$, even in the infinite
volume limt; therefore the same procedure should
work in two dimensions, corroborating the standard
result: asymptotic freedom for $N\ge 3$. The fact that
the $M\to\infty$ limit of the linear $\sigma$ model (performed
here), the known $N = \infty$ solutions, and Wilson's renormalization
group (ref.\ [1]) lead to the same answer
gives one yet more confidence in ordinary perturbation
theory, as applied to the $O(N)$ nonlinear $\sigma$ models.

\bigskip
I am indebted to John Preskill for suggesting this
problem and for many stimulating discussions, and
also to J.\ Feng and B.\ Warr for the latter. Lately
I learned that H.D.\ Politzer and G.\ Siopsis were carrying
out related investigations [10]. I thank them
for helpful exchanges of ideas.

\bibliographystyle{elsarticle-num}



\end{document}